\documentclass[onecolumn,secnumarabic,floatfix,amssymb,amsmath,nofootinbib,tightenlines,showpacs,superscriptaddress,aps,prb]{revtex4}
\usepackage{amsfonts}
\usepackage[colorlinks=true,linkcolor=blue,pagecolor=blue,filecolor=blue,menucolor=yellow,urlcolor=blue,citecolor=blue,anchorcolor=blue]{hyperref}%
\usepackage{graphicx}
\usepackage{dcolumn}
\usepackage{times}

\begin{document}
\title{Energy of Stable Half-Quantum Vortex in Equal-Spin-Pairing}

\author{Shokouh Haghdani}
\email{shokouhhaghdani@gmail.com}
\affiliation{Department of
Physics, Faculty of Sciences, University of Isfahan, 81744 Isfahan,
Iran}
\author{Mohammad Ali Shahzamanian}
\affiliation{Department of Physics, Faculty of Sciences, University
of Isfahan, 81744 Isfahan, Iran}

\begin{abstract}
In the triplet equal-spin-pairing states of both $^{3}He-A$ phase
and $Sr_{2}RuO_{4}$  superconductor, existence of Half-Quantum
Vortices $(HQVs)$ are possible. The vortices carry half-integer
multiples of magnetic quantum flux $\Phi_{0}=hc/2e$. To obtain
equilibrium condition for such systems, one has to take into account
not only weak interaction energy but also effects of Landau Fermi
liquid. Our method is based on the explanation  of the $HQV$ in
terms of a $BCS-$like wave function with a spin-dependent boots. We
have considered $\ell=2$ order effects of the Landau Fermi liquid.
We have shown that the effects of Landau Fermi liquid interaction
with $\ell=2$ are negligible. In stable $HQV$, an effective Zeeman
field exists. In the thermodynamic stability state, the effective
Zeeman field produces a non-zero spin polarization in addition to
the polarization of external magnetic field.
\end{abstract}

\maketitle

\section{Introduction}
One view to the liquid phase of $^{3}He$  contains in normal and
superfluid parts which at $3\times10^{-3}$ Kelvin the superfluid
part starts to be occurred mostly in triplet pairing
\cite{Leggeett1,Vollhardt}. The spin triplet pairing in the
superconductor compound $Sr_{2}RuO_{4}$  is observed experimentally
bellow $1.5$  Kelvin \cite{Mackenzie}.
\\The triplet pairing contains particles with the same spin directions that
leads to spin current. The spin current leads to interesting and
important phenomena like  $HQVs$ in equal-spin-paring  $(ESP)$.
Unlike common vortices, the half-quantum vortices contain
half-integer multiplications of the flux quantum $\Phi_{0}=hc/2e$ .
The origins of vortices in  type-II superconductors and $^{3}He$ or
$^{4}He$ superfluid are different. In the former case, the vortices
are appeared in the presence of external magnetic field while
appearance of the vortices in the latter cases is caused by the
rotation of the vessel which $^{3}He$ or $^{4}He$ is contained.
External magnetic field influences the vortices in $^{3}He$ too and
can generate half-quantum vortices.
\\Vakaryuk and Leggett\cite{Vakaryuk} have shown that in
equal-spin-pairing state, the stability condition of half-quantum
vortices is obtained when strong interactions also taken into
account. The general method for calculating the strong interactions
are presented by the Landau Fermi liquid theory. They have
considered only $\ell=1$ term in this interaction. In this work,
$\ell=2$ term is accounted and it is found that term with $\ell=2$
is very small compared to $\ell=0,1$ terms.

\section{Theoretical approach}
In the $ESP$ state of a spin triplet condensate, the spin of
particles in the Cooper pair is either aligned (up) or antialigned
(down) with a common direction in the space which is called $ESP$
axis\cite{Leggeett1,Vollhardt}. Therefore Cooper pairs may be
described via a linear superposition of states
$|\uparrow\uparrow\rangle$ and $|\downarrow\downarrow\rangle$. The
pairs are condensed in the same orbital states, we show them by
$\varphi_{\uparrow}$ and $\varphi_{\downarrow}$. The many-body wave
function which describes a system of $N/2$ pairs by
$\varphi_{\uparrow}$ and $\varphi_{\downarrow}$ can be written as
\begin{equation}\label{1}
\Psi_{ESP}=\emph{A}\{[\varphi_{\uparrow}(\textbf{r}_{1},\textbf{r}_{2})|\uparrow\uparrow\rangle+\varphi_{\downarrow}(\textbf{r}_{1},\textbf{r}_{2})|\downarrow\downarrow\rangle]...[\varphi_{\uparrow}(\textbf{r}_{N-1},\textbf{r}_{N})|\uparrow\uparrow\rangle+\varphi_{\downarrow}(\textbf{r}_{N-1},\textbf{r}_{N})|\downarrow\downarrow\rangle]\},
\end{equation}
where $\emph{A}$ is the antisymmetrization operator with respect to
particles coordinates $\textbf{r}_{i}$ and spins. Here, for
simplicity it will be assumed that the  $ESP$ state is neutral.
Annular geometry with the radius $R$ and the wall thickness $d$ with
$d/R\ll1$ is considered, to prevent complications connected to the
presence of the core of vortex. Therefore at  zero temperature, the
$HQV$ state of the condensate can be described via \cite{Vakaryuk}
\begin{equation}\label{2}
\Psi_{HQV}=\exp\{\frac{i\ell_{\uparrow}}{2}\sum_{i=\uparrow}\theta_{i}+\frac
{i\ell_{\downarrow}}{2}\sum_{i=\downarrow}\theta_{i}\}\Psi_{ESP},
\end{equation}
here $\theta_{i}$ is the azimuthal coordinate of the $i$th particle
on the annulus and the spin axis is along the symmetry axis of the
annulus. The integer $\ell_{\sigma}$ denotes the angular momentum of
$\sigma$, the component of pair wave function.\\
The thermal stability of system is obtained via minimization of
energy of the system by using the wave function in Eq. (\ref{2}). In
fact one  needs full version of original Hamiltonian essentially. In
the simplest case, one may consider $BCS$ Hamiltonian,
$\emph{H}_{BCS}$, along with spin triplet pairing term; however the
mentioned Hamiltonian $\emph{H}$ is unable to provide the
thermodynamic stability of the $HQV$. In such systems, the stability
condition is obtained when strong interactions taken into account.
The general method for calculation the strong interactions are
presented by the Landau Fermi liquid theory. Although the method was
used for investigating the normal metals, the generalized version of
the method is used for studying superconductor and superfluid
\cite{Leggeett1}. Thus Hamiltonian of the system involves two parts;
1)  $BCS$ Hamiltonian and 2) Landau Fermi liquid effects
,$\emph{H}_{FL}$, then $ \emph{H}=\emph{H}_{BCS}+\emph{H}_{FL}$.
First, the expectation value of the weak coupling Hamiltonian with
the state Eq. (\ref{2}) is calculated. In this case, one can write
it as a sum of three terms with different physical sources:
\begin{equation}\label{3}
E_{BCS}=E_{0}+E_{S}+T.
\end{equation}
where $E_{0}$ is the energy contribution originated from the freedom
internal degrees of Cooper pairs. This contribution is independent
on the center of mass motion of the Cooper pairs i.e., on quantum
numbers $\ell_{\uparrow}$ and $\ell_{\downarrow}$ and the magnetic
field magnitude for the annulus radius $R$ where is much larger than
the $BCS$ coherence length $\xi_{0}$ \cite{Vakaryuk1}. In this work
we assume a large enough annuls so that this term is ignorable.\\
The second term in the Eq. (\ref{3}) is the energy of spin
polarization of the system. We assume $N_{\sigma}$ as the particles
number with spin projection $\sigma$. one can define $S$ as a
projection of the total number spin polarization on the symmetry
axis as $S\equiv(N_{\uparrow}-N_{\downarrow})/2$. Therefor, spin
polarization energy is obtained as
\begin{equation}\label{4}
E_{S}=\frac{(g_{S}\mu_{B}S)^{2}}{2\chi_{ESP}}-g_{S}\mu_{B}\textbf{B.S},
\end{equation}
where $g_{S}$ is the gyromagnetic ratio for particles and
$\chi_{ESP}$ is the $ESP$ state spin susceptibility. It is important
that the spin polarization $S$, is a variational parameter and the
actual value of $S$ is obtained by minimization of the energy.\\
The third term in the Eq. (\ref{3}) is the kinetic energy of the
currents circulating in the system. We introduce the new parameters
as \cite{Vakaryuk}:
\begin{equation}\label{5}
\ell_{s\Phi}\equiv\frac{\ell_{\uparrow}+\ell_{\downarrow}}{2}-\frac{\Phi}{\Phi_{0}},
\ell_{sp}\equiv\frac{\ell_{\uparrow}-\ell_{\downarrow}}{2},
\end{equation}
where $\Phi$ is the total flux through the annulus. By using the
above introduced parameters, $T$ takes the following form:
\begin{equation}\label{6}
T=\frac{\hbar^{2}}{8m^{*}R^{2}}\{(\ell^{2}_{s\Phi}+\ell^{2}_{sp})N+4\ell_{sp}\ell_{s\Phi}\},
\end{equation}
where $m^{*}$ is the effective mass of particles containing the
Fermi liquid corrections. It is related to the bare mass of
particles $m$ by the usual relation of Fermi liquid theory as
$m^{*}=m(1+F_{1}/3)$ \cite{Leggeett1,Vollhardt}. In Eq. (\ref{6})
the first term in the brackets is constant. Because it is
proportional to the total number of particles $N\equiv
N_{\uparrow}+N_{\downarrow}$ and given values of $\ell_{s\Phi}$ and
$\ell_{sp}$. The second term generates an effective Zeeman field in
the $HQV$ state due to proportional to the spin polarization $S$.
The magnitude of this field and hence the value of the thermal
equilibrium spin polarization should be
obtain by energy minimization\cite{Vakaryuk}.\\
For making an $HQV$ stable, the strong coupling effects should be
considered \cite{Leggeett1,Leggeett2}. Also it needs to be accounted
the change of Fermi liquid energy $E_{FL}$ due to the presence of
spin and momentum currents in the $HQV$ state. These currents are
created by the spin dependent boost of Eq. (\ref{2}). We use the
standard formalism of Fermi liquid theory and investigate $\ell=2$
order effects of Landau Fermi liquid. By considering the mentioned
assumptions, we finally reach at the lengthly expression as follow:
\begin{eqnarray}\label{7}
\nonumber
E_{FL}=&&\frac{1}{2}(\frac{dn}{d\varepsilon})^{-1}Z_{0}S^{2}+\\&&\nonumber\frac{N^{-1}\hbar^{2}}{8m^{*}R^{2}}\frac{1}{3}\{(\ell^{2}_{s\Phi}F_{1}+\ell_{sp}^{2}\frac{Z_{1}}{4})N^{2}+4(\ell^{2}_{sp}F_{1}+\ell_{s\Phi}^{2}\frac{Z_{1}}{4})S^{2}
 +4\ell_{s\Phi}\ell_{sp}(F_{1}+\frac{Z_{1}}{4})SN \}\\&&
\nonumber+\frac{N^{-1}\hbar^{4}}{64m^{*}p^{2}_{F}R^{4}}\{[(F_{2}+\frac{Z_{2}}{4})(\ell_{s\Phi}^{2}+\ell_{sp}^{2})^{2}+\frac{Z_{2}}{2}\ell_{s\Phi}^{2}\ell_{sp}^{2}]N^{2}+[(16F_{2}+2Z_{2})\ell_{s\Phi}^{2}\ell_{sp}^{2}]\times\\&&S^{2}+[(8F_{2}+2Z_{2})
\ell_{s\Phi}\ell_{sp} (\ell_{s\Phi}^{2}+\ell^{2}_{sp})]
SN\}-\frac{N^{-1}p_{F}^{2}}{12m^{*}}(F_{2}N^{2}+Z_{2}S^{2}),
\end{eqnarray}
where $(dn/d\varepsilon)$ is the density of states at the Fermi
surface and $Z_{0}, Z_{1}, Z_{2}, F_{0}$ and $ F_{1}$ are Landau
parameters. The first term in the above equation, proportional to
$Z_{0}$, is the energy cost generated by a spin polarization and the
rest explains Fermi liquid corrections caused by the presence of
currents.\\
Now we obtain the equilibrium spin polarization in the $HQV$ state
by minimizing the total energy $E_{BCS}+E_{FL}$ with respect to $S$.
Then $S$ can be written as
\begin{equation}\label{8}
S=(g_{s}\mu_{B})^{-1}\chi B^{'},
\end{equation}
where $\chi$ is the spin susceptibility of the system. For
$^{3}He-A$ the value of $\chi$ is approximately $0.37$ times smaller
than the normal state susceptibility at low temperatures
\cite{Leggeett2}. We also find $\chi$ as:
\begin{eqnarray}\label{9}
\nonumber
\chi^{-1}=\frac{1}{\chi_{ESP}}+(\frac{d n}{d\varepsilon})^{-1}Z_{0}(g_{S}\mu_{B})^{-2}+\frac{N^{-1}\hbar^{2}}{m^{*}R^{2}(g_{S}\mu_{B})^{2}}(\ell_{sp}^{2}\frac{F_{1}}{3}+\ell_{s\Phi}^{2}\frac{Z_{1}}{12})\\
+\frac{N^{-1}\hbar^{4}}{16m^{*}p_{F}^{2}R^{4}(g_{S}\mu_{B})^{2}}\ell^{2}_{sp}\ell^{2}_{s\Phi}(8F_{2}+Z_{2})-\frac{N^{-1}p^{2}_{F}Z_{2}}{6m^{*}(g_{S}\mu_{B})^{2}}.
\end{eqnarray}
The Zeeman field $B^{'}$ is involved two parts: 1) the external
Zeeman field $B$ and 2) the effective Zeeman field $B_{eff}$:
\begin{equation}\label{10}
B^{'}=B+B_{eff},
\end{equation}
The effective Zeeman field $B_{eff}$ is due to the present of spin
currents:
\begin{equation}\label{11}
B_{eff}=-\frac{\hbar^{2}(g_{S}\mu_{B})^{-1}}{2m^{*}R^{2}}\ell_{sp}\ell_{s\Phi}\{1+\frac{F_{1}}{3}+\frac{Z_{1}}{12}+\frac{\hbar^{2}}{16p^{2}_{F}R^{2}}(4F_{2}+Z_{2})(\ell^{2}_{s\Phi}+\ell^{2}_{sp})\}.
\end{equation}
The effective Zeeman field is a periodic function of the total
external magnetic flux $\Phi$ with a period equal to $\Phi_{0}$. The
sign of the effective field is altered when the total magnetic field
is equal to half-integer values of the flux quantum\cite{Vakaryuk}.\\
Finally, the energy of the system $E\equiv\langle H\rangle$ can be
obtained by inserting the value of $S$, according to Eq. (\ref{8}) -
Eq. (\ref{11}), to the relation of the total energy and ignoring the
internal energy contribution. Then one can write:
\begin{equation}\label{12}
E=-\frac{1}{2}\chi
B^{'}+\frac{\hbar^{2}N}{8mR^{2}}\{\ell^{2}_{s\Phi}+\ell^{2}_{sp}\frac
{1+\frac{Z_{1}}{12}}{1+\frac{F_{1}}{3}}
\}+\frac{\hbar^{4}N}{64mp^{2}_{F}R^{4}}\{(\ell^{2}_{s\Phi}+\ell^{2}_{sp})^{2}\frac{F_{2}+\frac{Z_{2}}{4}}{1+\frac{F_{1}}{3}}+\ell^{2}_{s\Phi}
\ell^{2}_{sp}\frac{Z_{2}}{1+\frac{F_{1}}{3}}\}.
\end{equation}
The contribution of the spin polarization to the energy  ( first
term of Eq. (\ref{12})) is small for reasonable values of the
external magnetic field. The third term of the equation is related
to $\ell=2$ which compare to the second term is order of
$\hbar^2/2mR^2\varepsilon_F$. Evidently, one can ignore them for
analyzing the stability of $HQV$s. The stability region of the
$HQV$s depends on the $(1+Z_1/12)/(1+F_1/3)$ that is, the ratio of
superfluid spin density to superfluid density $\rho_{sp}/\rho_s$
\cite{Leggeett1}. The criteria of stability is directly obtained by
minimizing of Eq. (\ref{12}) and leads to $\rho_{sp}/\rho_s\prec 1$
\cite{Volovik}. In the $^3He-A$, the ratio is less than unity for
all temperatures bellow critical temperature and then $HQV$s are
possible for a large limit of the phases diagram\cite{Volovik}.
\section{Conclusions}
In the equal-spin pairing condensation of $HQV$ an effective Zeeman
field $B_{eff}$ exists. In the thermodynamic stability state, the
effective Zeeman field produces a non zero spin polarization in
addition to the polarization of external magnetic field $B$. The
thermal stability of system is obtained via minimization of energy
of the system by using the wave function and the total Hamiltonian.
The best Hamiltonian of the system involves two parts; 1) $BCS$
Hamiltonian and 2) Landau Fermi liquid Hamiltonian. The effects of
$\ell=2$ terms of  Landau Fermi liquid have been considered in this
paper. The third term in Eq. (\ref{12}) contains Landau parameters
$Z_{2}$ and $F_{2}$, which compare to the second term is order of
$\hbar^2/2mR^2\varepsilon_F$. This quantity with $R=0.1$ micron
takes approximately $10^{-7}$. Therefore, one can omit this term in
Eq. (\ref{12}). Consequently for the stability condition of $HQV$s,
it is sufficient that the second term of Eq. (\ref{12}) be
considered and $\frac{\rho_{sp}}{\rho_{s}}<1$ is obtained
 for stability condition.


\end{document}